# Evolution of an oxygen NEXAFS transition in the upper Hubbard band in α-Fe$_2$O$_3$ upon electrochemical oxidation

Debajeet K. Bora[1,2], Artur Braun[1*], Selma Erat[1,3], Romy Löhnert[1,4], Kevin Sivula[5], Jörg Töpfer[4], Michael Grätzel[5], Thomas Graule[1,6], Edwin Constable[2]

[1]*Laboratory for High Performance Ceramics*
*Empa. Swiss Federal Laboratories for Materials Science and Technology*
*CH-8600 Dübendorf, Switzerland*

[2]*Department of Chemistry, University of Basel*
*CH-4056 Basel, Switzerland*

[3]*ETH Zürich, Department of Non-metallic Materials*
*CH-8093 Zürich, Switzerland*

[4]*Department of SciTec*
*University of Applied Sciences Jena*
*D-07745 Jena, Germany*

[5]*Laboratory for Photonics and Interfaces*
*Ecole Polytechnique Federale de Lausanne*
*CH-1015 Lausanne*

[6]*Technische Universität Bergakademie Freiberg*
*D-09596 Freiberg, Germany*

---

* Corresponding author: Phone +41 44 823 4850, Fax: +41 44 823 4150,
email artur.braun@alumni.ethz.ch

ABSTRACT

Electrochemical oxidation of hematite (α-Fe$_2$O$_3$) nano-particulate films at 600 mV vs. Ag$^+$/AgCl reference in KOH electrolyte forms a species at the hematite surface which causes a new transition in the upper Hubbard band between the Fe(3d)-O(2p) state region and the Fe(4sp)-O(2p) region, as evidenced by oxygen near edge x-ray absorption fine structure (NEXAFS) spectra. The electrochemical origin of this transition suggests that it is related with a surface state. This transition, not known for pristine α-Fe$_2$O$_3$ is at about the same x-ray energy, where pristine 1% Si doped Si:Fe$_2$O$_3$ has such transition. Occurrence of this state coincides with the onset of an oxidative dark current wave at around 535 mV – a potential range, where the tunneling exchange current has been previously reported to increase by three orders of magnitude with the valence band and the transfer coefficient by a factor of 10. Oxidation to only 200 mV does not form such extra NEXAFS feature, supporting that a critical electrochemical potential between 200 and 600 mV is necessary to change the electronic structure of the iron oxide at the surface. Decrease of the surface roughness, as suggested by visual inspection, profilometry and x-ray reflectivity, points to faceting as potential structural origin of the surface state.

# Introduction

Hematite, α-Fe$_2$O$_3$ is the most common iron oxide mineral in nature, and a detailed understanding of its electronic structure and transport properties is of particular interest for its functionality in device applications. Particularly for photoelectrochemical applications, α-Fe$_2$O$_3$ is attractive as a photoanode due to its suitable durability, abundance, and valence band edge position with respect to water oxidation potential [1-5]. Hematite is also regarded as the cheapest semiconductor that absorbs substantial amounts of visible light, and is therefore a candidate component of inexpensive, inorganic artificial photosynthesis systems for generating chemical fuels from sunlight [6]. Its electronic structure has been studied with x-ray and electron spectroscopy in its pristine state as well as after exposure to gases and, because of its importance to geosciences, exposure to water. Insofar, a cross fertilization between materials science and geological science has taken place in the last ten years. When hematite is used in a photoelectrochemical cell, the interaction of aqueous electrolyte with hematite under the applied potential may have an influence on its electronic structure especially at the semiconductor liquid junction.

The crystallographic structure of the [0001] hematite surfaces in aqueous media has been studied by scanning tunneling microscopy [7]. Particularly soft x-ray spectroscopy at the oxygen K-edge has proven useful for the analysis of the electronic structure [8,9,10,11,12,13]. The O NEXAFS spectrum of hematite originates from transition from the O (1s) core orbitals to the empty Fe (3d)-Fe(4s)- and Fe (4p)- like bands or molecular orbitals that have some O (2p) character. The surface oxidation phase on pyrite FeS$_2$ reacted in aqueous electrolytes at pH 2-10 and with air under ambient condition was studied using synchrotron–based O K-shell near edge x-ray absorption fine structure (NEXAFS) spectroscopy [14,15]. NEXAFS spectroscopy at the O K-edge was also utilized to reveal band edge electronic structure of bulk and nanoscale hematite [16]. The band gap widening in nanoscale iron oxide is potentially important for its application in solar cells [17,18]. Thin film of single crystal iron oxides, Fe$_x$O (111), Fe$_3$O$_4$ (111) and α-Fe$_2$O$_3$ (001) prepared by oxidizing the Fe films evaporated on a Pt (111) surface, have been studied recently using NEXAFS [19]. NEXAFS studies of the electronic structure and chemistry of iron based metal oxide nanostructured materials showed the correlation between electronic structure and surface chemistry [10].

The motivation of our work is to study the influence of electrochemical oxidation of hematite on its electronic structure, because such kind of oxidation is an operational step in the functionality of photoelectrochemical cells. For this we have carried out O K-edge NEXAFS spectroscopy of hematite films with different thicknesses cycled under applied bias of 200 mV and 600 mV in 1M KOH. The spectra are compared with spectra obtained from Si doped hematite films prepared by atmospheric pressure chemical vapor deposition. To the best of our knowledge, hematite before and after electrochemical treatment have not yet been studied with NEXAFS spectroscopy.

## Experimental Section

**Reagents and Materials.** Hematite α-$Fe_2O_3$ thin films were synthesized by dip coating of a precursor complex on a fluorinated tin oxide (FTO, TEC-8 Pilkington from Hartford Glass, MA) coated glass substrate, followed by annealing at 500°C for 2 h in air. The precursor complex was synthesized by heating a mixture of $Fe(NO_3)_3 \cdot 9H_2O$ (28.0 g) and oleic acid (17.0 g) to 70°C to give a homogenous liquid phase. This homogeneous mixture was then heated at 125°C for 90 minutes to give a reddish brown viscous mass which was then cooled to room temperature, left for 24 h and subsequently treated with 80 ml of tetrahydrofuran. The resultant solution was stirred with a glass rod for 30 minutes and the powdery precipitate separated from the solution by centrifugation (5000 rpm) for 2-3 min. After centrifugation, the supernatant (precursor complex) was recovered for the dip coating of the film. FTO was used as a substrate. By repeated dipping and annealing at 500°C for 30 minutes for each layer, films with layer thicknesses from 1 to 10 layers were obtained. This deposition technique is reproducible and allows making porous films with well defined thickness by repeated dip coating. The thickness of a number of films was determined with a stylus profilometer (Ambios XP-100). 4 dip coated layers resulted in an approximately 600 nm thick film with optimized photoelectrochemical properties. Phase purity was confirmed by x-ray diffraction. Thicker films had overall larger crystallite sizes, as evidenced by the evolution of widths of Bragg reflections. 1 at.% Si doped α-$Fe_2O_3$ films were deposited on the same FTO substrates with atmospheric pressure chemical vapor deposition (APCVD) after a previously described protocol [20]. Photocurrent and dark current were measured using a spectro-electrochemical cell containing 1 M KOH as electrolyte, $Ag^+$/AgCl reference electrode and a Pt counter electrode. Chronoamperometry was applied at 200 mV for 2 hours to oxidize the films electrochemically. Near edge x-ray absorption fine structure (NEXAFS) spectra were recorded at the undulator beamline UG-56 at BESSY [21], in an UHV recipient with 2 x $10^{-10}$ mTorr base pressure or lower. The resolution of this beamline is 0.05 eV (80000 at 64eV) at the soft energy range for oxygen K-edge and Fe L-edges. X-ray reflectometry data were recorded with a Siemens D5000 diffractometer with Cu Kα1 radiation in θ/2θ-configuration in a range of $0.07° \leq 2θ \leq 2°$ with a scan speed of 0.5s per 0.002° step.

**Results and Discussion**

The evolution of the film thickness upon repeated dip coating and subsequent annealing is shown in Figure 1–a. The linear increase as a function of coated layers, with an average thickness of 164±3 nm per layer as determined by least square regression, demonstrates reproducibility of the deposition method. Note, that the first layer in a 10 layer thick film has been exposed 10 times to 500°C for 30 minutes, whereas the top layer has been exposed only once. The film should thus have graded structure across the thickness. This is indicated by the average crystallite size, which ranges from 30 nm for the 1 layer film to 80 nm for the 10 layer film according to x-ray diffraction. Phase purity of hematite has been confirmed by x-ray diffraction [22].

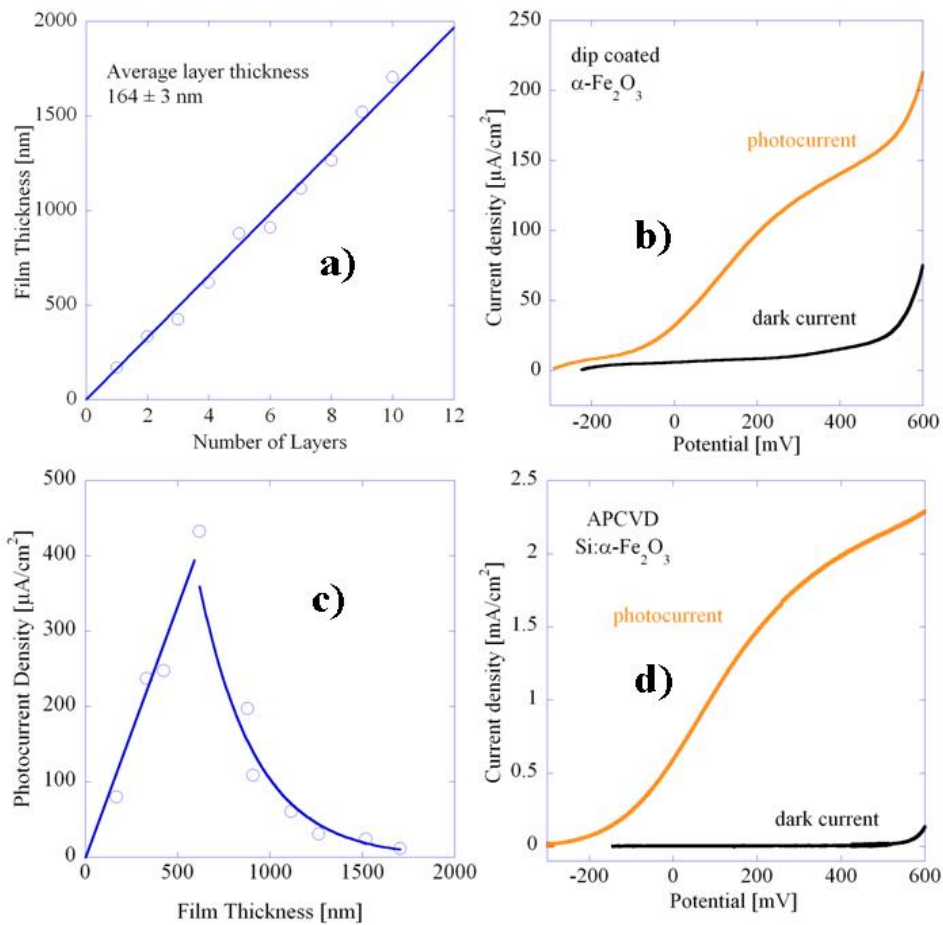

**Figure 1:** a) – Film thickness as a function of layers with linear least square fit; b) – photocurrent and dark current of 1 layer film obtained by dip coating; c) - photocurrent of dip coated films as a function of film thickness, measured at 600 mV; d) – photocurrent and dark current of Si-doped α-$Fe_2O_3$ obtained by APCVD.

The current/voltage diagram in Figure 1-b shows a strong onset of photocurrent for a 1 layer film with 164 nm thickness at around 0mV, whereas the dark current maintains a flat slope of around 15µA/cm$^2$ until 500 mV. At the same potential also the photocurrent has a strong additional onset of an oxidative current wave. For films with 1 to 9 layers, the current was determined at 600 mV and plotted versus the film thickness, see Figure 1-c. The photocurrent is obviously dependent on the thickness of the film. The total current measured at 600 mV includes a substantial dark current, which has not been subtracted in Figure 1-c. For films with up to 4 layers, the photocurrent increases linearly. The maximum photocurrent was found for a film with 4 layers and a thickness of 650 nm, i.e. 430μA/cm$^2$. For films with more than 4 layers, the photocurrent decreases with a profile that can be modeled with an exponential. Thus, the photocurrent can be optimized in terms of film thickness. For comparison, Figure 1-d shows dark current and photocurrent of a Si-doped hematite film at about the same potential like the one shown in Figure 1-b.

Figure 2 shows the oxygen NEXAFS spectra of four differently treated α-Fe$_2$O$_3$ films with 4 layers deposited on FTO glass. The spectra have been normalized to the intensity tail at energies larger than 550 eV. One spectrum corresponds to the pristine film, one spectrum to such film exposed to KOH electrolyte, and two films electrochemically oxidized under dark and under light condition from 0 mV to 200 mV. The dark/light experiment was carried out because hematite has a charge-transfer band gap (about 2.2 eV) smaller than the cutoff of solar radiation in the troposphere (about 4.3 eV) and is thus able to participate in photochemical reactions [23]. At first glance we see no major differences in these four spectra. Particularly all resonances are virtually at the same energy positions. The spectra show a well developed double peak in the pre-edge region at 530 eV, see upper right inset in Figure 2, which originates from transitions of hybridized Fe(3d)-O(2p) states with $t_{2g}$ and $e_g$ orbital symmetry [8,9,12,13]. Oxygen edge profiles are sensitive to the local bonding and symmetry properties of the excited oxygen. The features of the pre peak are governed by the 3d components in the hybridized unoccupied pd wave functions [13]. A potential difference is the intensity at around 533-534 eV, the range between the Fe(3d)-O(2p) doublet and the Fe(4sp)-O(2p) resonances. It is low for the pristine sample (red color spectrum), whereas the spectra of the samples exposed to KOH and oxidized to 200 mV have a higher intensity in this energy region, as well as in the region where the doublet is found.

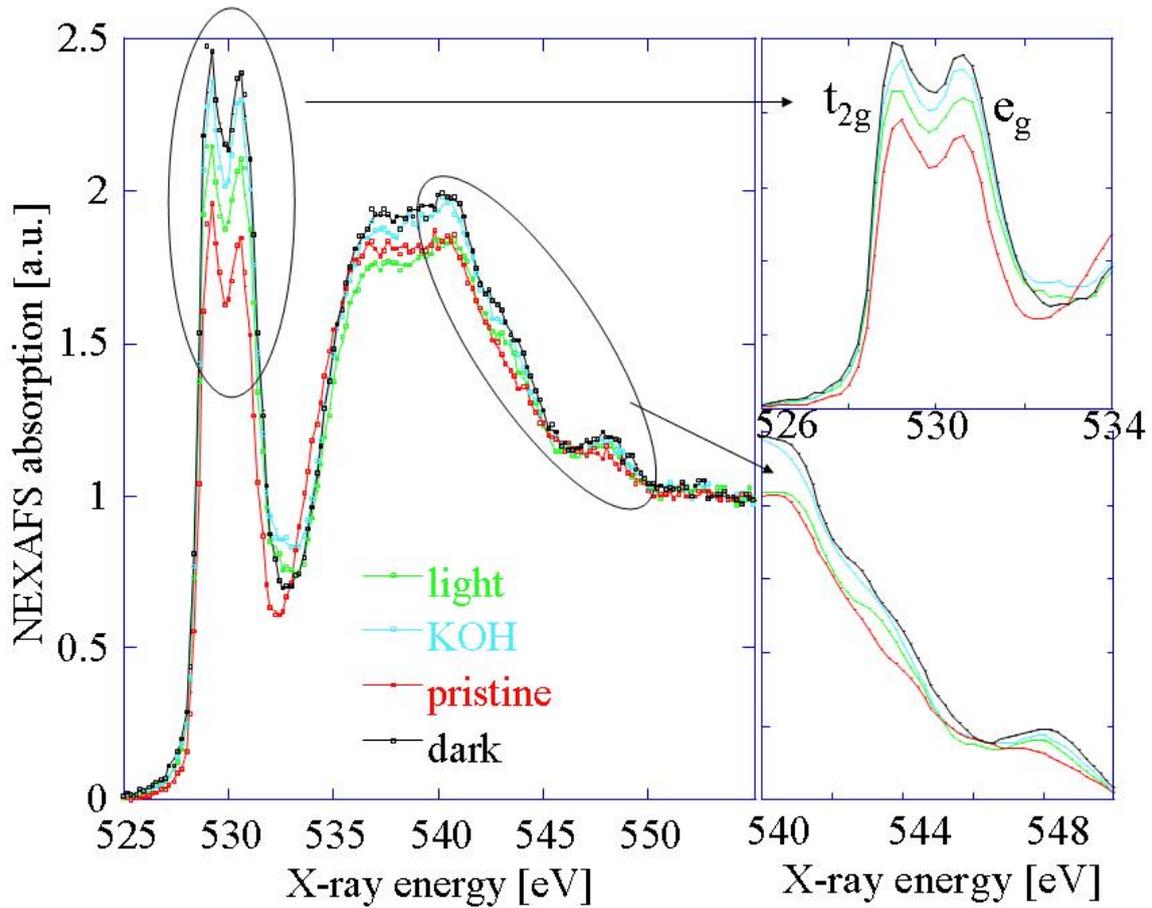

**Figure 2**: Oxygen NEXAFS spectra of pristine α-$Fe_2O_3$ film, film exposed to KOH, and light and dark treated film from 0 mV to 200 mV in KOH for 2 hours.

Close inspection and comparison of all four spectra in Figure 2 suggests that upon KOH exposure, a weak broad structure readily present in the pristine sample at around 542 eV to 544 eV evolves a little more upon contact with KOH and oxidation to 200 mV. The spectral range for this structure is shown in the magnified lower inset in Figure 2. In addition it appears that the spectra of the samples exposed to KOH have a slight noticeable shift in the energy range at 542 eV towards higher energy. Another peculiarity is that the pre-edge peaks of the KOH treated samples are relatively larger than those of the pristine sample.

We find more peculiar details in the films that were oxidized at a voltage of 600 mV. These films reached the potential of 600 mV only during the photocurrent and dark current measurement, i.e.

only for several seconds. Figure 3 shows the oxygen NEXAFS spectra of a pristine 4 layer film, and films with 1 layer and l0 layers, both of which were oxidized to 600 mV, and the oxygen spectrum of a clean FTO glass for reference.

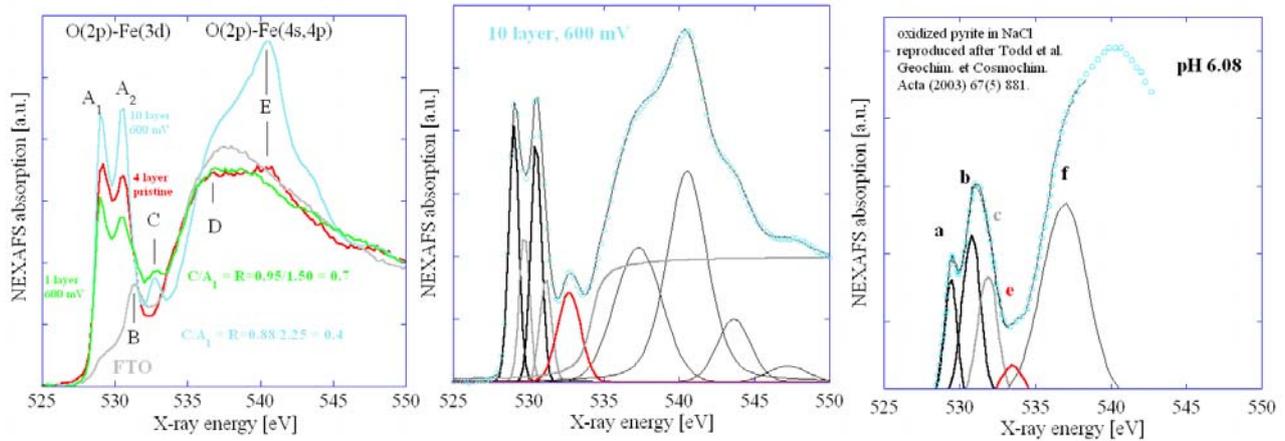

**Figure 3**: Left - Oxygen NEXAFS spectra of FTO, pristine $Fe_2O_3$ film 4 layers thick (red spectrum), and 1 layer (green) and 10 layer (blue) thin films oxidized to 600 mV in KOH. Middle – deconvolution of 10 layer film spectrum into Voigt functions and arctan function. The extra peak is plotted in red. Right – Oxygen NEXAFS spectrum of oxidized pyrite, as reproduced from Todd et al., ref. [8].

The spectrum of the FTO glass has only minor intensity in the range where the hematite has a typical strongly developed doublet – around 530 eV. FTO has a relatively strong resonance at 531.4 eV. These spectral details are important in order to judge which details in the spectra originate from $Fe_2O_3$, and which are from the FTO underneath. Therefore it is also important to know about the information depth and x-ray attenuation depth of the x-rays in the $Fe_2O_3$ sample. For $Fe_2O_3$ with an assumed density of 5.3 g/cm$^3$, the theoretical attenuation depth for x-ray energies of around 530 eV is around 600 nm, this is, the x-ray intensity of 530 eV electrons has decayed to 1/e of its original intensity [24]. According to Figure 1, the 1 layer film has a thickness of around 164 nm. Therefore, the oxygen spectrum of the 1 layer film is significantly contaminated with spectral intensity from the FTO underneath. This is clearly visible at around 531.75 eV for the 1 layer film, where a shoulder is observed which coincides with the aforementioned peak of the FTO sample in the same energy range. The films with 4 layers and 10 layers, 650 nm and 1650 nm thickness, respectively, do not show such shoulder.

We now take a closer look at the energy range of around 532.4 eV, Figure 3 left panel, where the pristine $Fe_2O_3$ and the FTO spectrum have a relative intensity minimum. We notice that the 10 layer film has a relatively strong transition (peak feature denoted C) in this region. Also, the 1 layer film has a transition in this region. This transition "C" at around 532.7 eV has not been observed previously in spectra of hematite. X-ray spectroscopy classifies hematite as a charge transfer or intermediate type insulator [8]. We believe that the extra peak C is due to a transition probably to the conduction band, i.e. to the upper Hubbard band with mainly Fe(3d) character. A potential scenario is transitions to on-site Fe empty 4p states hybridized to empty 3d states of next-to-nearest neighbor Fe atoms. Possibly, the C peak is thus determined by inter-site Fe 4p-3d mixing via strong Fe 3d-O 2p hybridization and thus a direct probe of the upper Hubbard band in this highly oxidized form of Fe; compare ref. [25]. The pristine $Fe_2O_3$, the $Fe_2O_3$ films exposed to KOH, and $Fe_2O_3$ oxidized to 200 mV, as discussed before, do not show this C peak. Hence, the electrochemical oxidation to 600 mV is a clear condition for the formation of this transition. Little is known about the redox chemistry and higher redox states of Fe at the iron oxide solid/aqueous liquid interface [26]. A representative energy diagram of the α-$Fe_2O_3$/KOH junction is sketched in [27]. Cummings et al. hypothesize the presence of Fe with higher oxidation states, i.e. $Fe^{4+}$ at the iron oxide interface with neutral electrolytes, in addition to readily established $Fe^{4+}$ at alkaline electrolytes [26, and references therein].

The pre-edge doublet in the oxygen NEXAFS spectra of FeOOH has been considered as a convolution of a pair of $t_{2g}$-$e_g$ states from the oxygen in FeOOH O(2p)-Fe(3d) transitions, and a conjugate pair of the hydroxyl group, latter of which is slightly shifted towards higher energy [28]. Comparison with the literature spectrum of FeOOH suggests that the extra peak that we observe at 532.7 eV is not commensurate with the peak labelled "d" in Figure 2-b) in ref. [14]. However,

Figure 6-a) in [14] shows three spectra obtained at high pH with a peak labeled "e", which remains unassigned in [14], in the same relative energy position like our extra peak at 532.7 eV. Todd et al. speculate this peak cannot be assigned to a bound state, but could be a multiple scattering feature [15]. We have reproduced this spectrum in the right panel in Figure 3 and highlighted feature "e" by deconvolution of the relevant part of the spectrum into a series of Voigt functions. In a follow up paper, Todd et al. [15] assign a similar structure in oxidized chalcopyrite to an O(1s) in $OH^-$ -> O(2p)-H(1s) antibonding molecular orbital in $OH^-$, possibly from adsorbed water. We believe this peak is representative to a highly oxidized surface species, slightly more so than from FeOOH because it is found at a higher energy than the feature specific to FeOOH, and potentially resembles an electron hole feature which transfers from O(2p) to Fe(3d). Increased spectral weight in the range around 534 eV, this is where the NEXAFS spectra have a relative minimum intensity, is also found in a size dependent study on hematite nanoparticles with O(1s) NEXAFS spectroscopy [16]. The spectra of nanoparticles have in this relative intensity minimum a larger spectral weight, than bulk hematite (see Figure 5-b) in [15]). The spectrum of hematite exposed to KOH only, Figure 2, suggests a similar increased intensity in this energy range, though not to the extent that a distinct peak is formed.

Comparison of the spectra from the 1 layer and 10 layer films in Figure 3 suggests that the 1 layer film has a relatively higher intensity of the extra transition C compared to the intensity of the doublet $A_1$, $A_2$, in contrast to the 10 layer film. This becomes clear in the relative peak height ratios between the newly formed peak C and the height of $A_1$ in the doublet. For the 1 layer film the relative ratio of the peak C is $R = C/A_1 = 0.95/1.50 = 0.7$, and for the 10 layer film it is $R = C/A_1 = 0.88/2.25 = 0.4$. One explanation is that the spectra which we compare here resemble not only the electronic structure of the surface but also to some extent the electronic structure of the bulk. We recall that the 10 layer film has larger hematite crystallites (average 80 nm) than the 1 layer film (average 30 nm) because of the extended exposure to 500°C. This implies that the 1 layer film with the relatively larger extra peak has relatively more surface contribution from that film than the 10 layer thick $Fe_2O_3$ film, where more of the information comprises the bulk properties. After all, the electrochemical oxidation in KOH should particularly impact the surface, and not the bulk. Impedance studies on electrochemical passivation of Fe metal [29, and references therein] point to an interconversion of p-type Fe(II) and n-type Fe(III) in the course of forming and dissolving a surface passive film. Specifically, within a potential range from -300 mV to +1000 mV an n-type Fe(III)-oxide layer is formed [29]. Moreover, between 500 mV and 700 mV, the tunneling exchange current increases by three orders of magnitude with the valence band, and the apparent tunneling current transfer coefficient decreases by a factor of ten [29]. From oxidation of pyrite it is known that at high pH iron(III) oxyhydroxide forms at the surface, and under most alkaline conditions the oxygen NEXAFS spectrum resembles that of goethite FeOOH [14]. When we compare feature E in the spectra at 540.5 eV (Figure 3), we notice its clear absence in the 1 layer film, which represents a larger surface-volume ratio than the thicker films and also has smaller crystallites. The 4 layer film has a noticeable intensity for feature E, and the 10 layer thick film has an overwhelmingly dominant E structure. We thus conclude the transition labeled E at 540.5 eV is a bulk hematite feature.

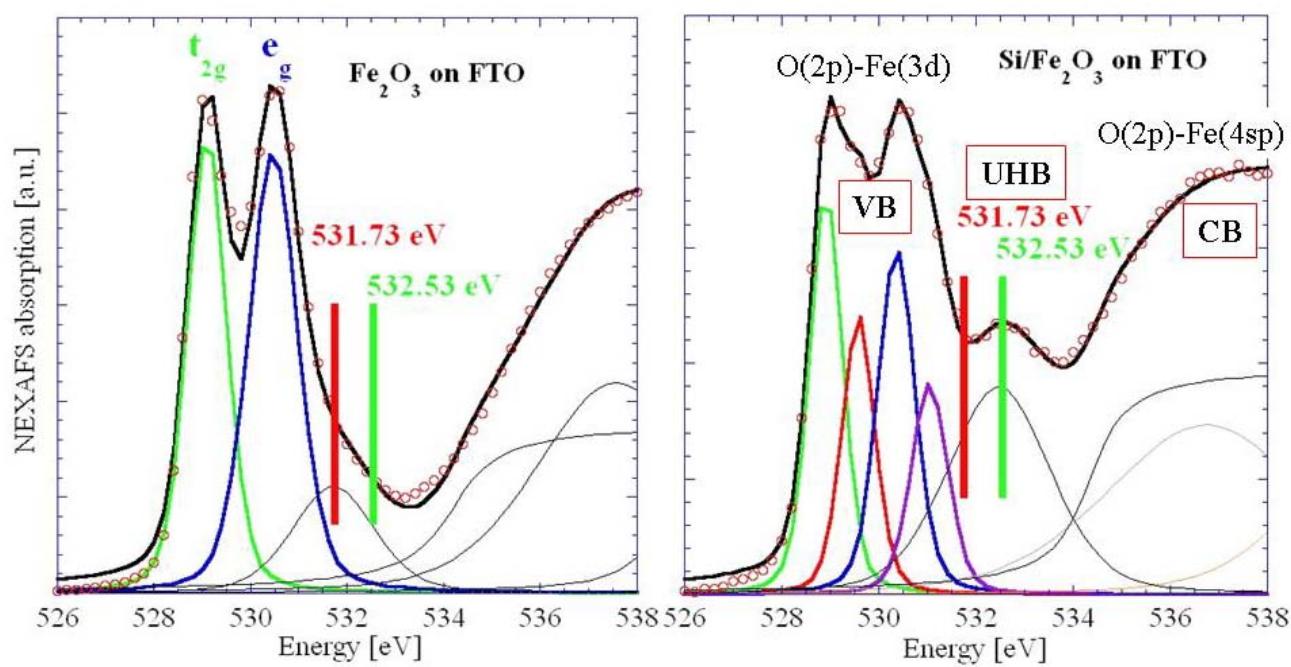

**Figure 4:** Oxygen NEXAFS spectra of α-Fe$_2$O$_3$ nanoparticle powder with 550°C heat treatment (left spectrum) and 1% Si-doped Fe$_2$O$_3$ pristine sample (right).

When we return to the $t_{2g}$-$e_g$ doublet features $A_1$ and $A_2$ at about 529 and 530 eV, we notice that the relative height of $A_1$ and $A_2$ is equal for the 10 layer film, whereas for the thinner films feature $A_1$ is larger than $A_2$. The likely reason for this is that the 10 layer film has been exposed 10 times to thermal treatment, whereas the 4 layer and 1 layer films have been exposed only 4 times and 1 time, respectively. Thus, the hematite phase in the thick film is crystallographically more developed, as was found by peak width analysis of the x-ray diffractograms, which manifests in the oxygen NEXAFS spectra pre-edge doublet, which has equal height in Fe six fold coordinated by oxygen [10]. Figure 4 shows the oxygen NEXAFS spectra of pristine α-$Fe_2O_3$ powder (i.e., no FTO underneath) and a 1% Si-doped α-$Fe_2O_3$ film synthesized by APCVD. Both spectra are reminiscent of α-$Fe_2O_3$ as far as the $t_{2g}$-$e_g$ doublet at 529 eV – 531 eV is concerned. Interestingly, the spectrum of the Si doped hematite shows an extra peak at around 532.4 eV - the same energy where in Figure 3 the feature C peak was found in the 600 mV oxidized samples - although the 1% Si-doped α-$Fe_2O_3$ film was not electrochemically treated. We have then cycled such film several times in KOH (not shown here), even to potentials exceeding 600 mV. Both spectra, the pristine and the electrochemically treated 1% Si-doped α-$Fe_2O_3$ film, show the feature C peak. The vertical bars (red and green at 531.7 and 532.5 eV) in the spectra in Figure 4 indicate the position of C, and a side band observed in the spectrum of the powder. Comparison shows clearly that the sideband of α-$Fe_2O_3$ does not coincide with our newly observed peak C, and both should not be mistaken for each other. Also other ferrous or ferric oxides do not show such extra peak in the oxygen NEXAFS spectra before or after exposure to water [30].

Peculiarities in the doublet structure of the APCVD 1% Si-doped $Fe_2O_3$ film spectra, particularly their width, made us believe that each peak in the doublet could be actually comprised of two peaks, so that the doublet shows actually four peaks in total. Awareness of potentially more transitions in the pre-edges of the oxygen spectra on doped oxides arises from experience with nitrogen doped $TiO_2$ [31]. Close inspection of the energy range from 531 eV to 534 eV shows that the pristine α-

$Fe_2O_3$ powder spectrum has a shoulder at around 531.7 eV which cannot be accounted for by the doublet in the deconvolution of the spectrum. We have therefore included an additional, broad Voigt function for its deconvolution at this energy.

The question naturally arises as to which influence has the electrochemical oxidation at 600 mV on the hematite surface, in contrast to oxidation to 200 mV, or merely the exposure to KOH. Does oxidation at 600 mV increase the hematite surface area? Upon visual inspection, the electrochemically oxidized films looked smoother than the pristine one. The Si-doped hematite films looked smoother even without any electrochemical treatment. This suggestion is quantitatively confirmed by profilometer data (not shown here), telling that larger spikes became reduced in size upon 2 hours oxidation at 600 mV. The x-ray reflectograms in Figure 5 also show that the reflectivity of the oxidized film exceeds that of the pristine film.

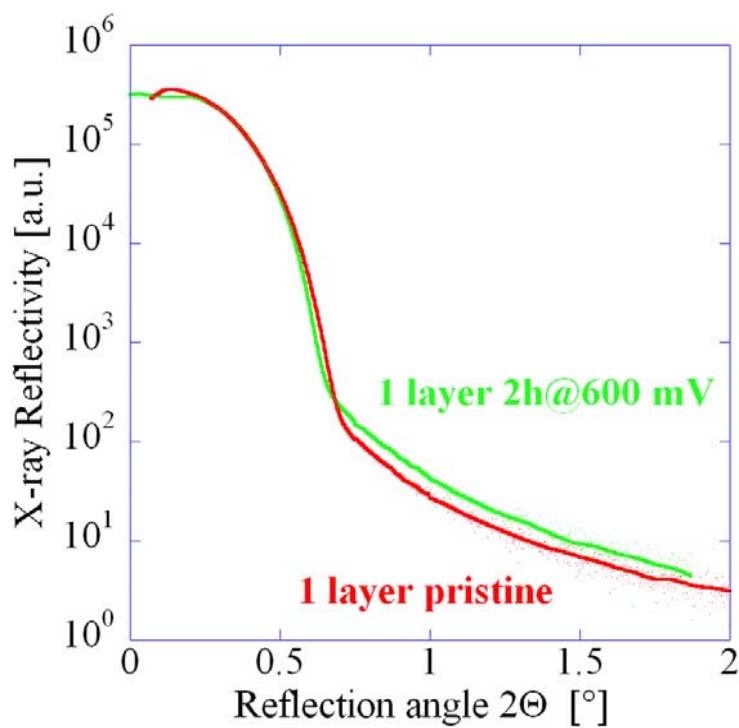

**Figure 5:** X-ray reflectometry data for the 1 layer α-$Fe_2O_3$ film before (pristine) and after electrochemical oxidation at 600 mV for 2 hours.

Hence, there is no or little support for a suggestion that the electrochemical oxidation to 600 mV would increase the surface roughness, at least not at the sensitivity scale for reflectometry. An alternative explanation is that the Si-doping and the electrochemical oxidation at 600 mV cause morphological changes in a way that a surface is reorganized, such as like formation of facets, which favor the formation of highly oxidized Fe species at the surface of the hematite. However, more elaborate surface structure studies are necessary in order to corroborate this point.

**Conclusion**

A new transition has been identified in the oxygen NEXAFS spectra on Si doped hematite and on electrochemically oxidized hematite. The energy position of this transition at around 532.7 eV relative to the top of the valence band suggests that this transition arises from O(2p) orbitals to the Hubbard band with a strong Fe(3d) character. While exposure to KOH seems to cause slight increase of spectral weight in this region, it needs a potential of around 600 mV to form this transition, the potential which reportedly causes the tunneling exchange current to increase by three orders of magnitude with the valence band, whereas a potential of 200 mV only is not sufficient for that. The electrochemical origin of this transition suggests that it is related with a surface state. In such alkaline environment, formation of highly oxidized Fe species is likely to occur, which may go along with electron hole states from the O(2p) orbitals. Without further experiments, a similar statement cannot be made for the Si doped hematite, which, interestingly, bears this electronic structure in absence of KOH or electrochemical potential. All films with this transition look visually smoother and have also, in the case of the 600 mV oxidized film, a higher x-ray reflectance, ruling out that increased surface roughness is a precondition for this peak. The fact that films with larger crystallites have relatively less spectral weight for this extra peak than those with smaller crystallites leads to the suggestion that surface faceting could be related with the extra transition to the Hubbard band. Support for this hypothesis comes from the suggestion that doping with Si acts as a hematite structure directing agent.

**Acknowledgement**

Funding by E.U. MIRG No. CT-2006-042095, Swiss NSF # 200021-116688 and 206021-121306. Swiss Federal Office of Energy project # 100411. We are indebted to Dr. A. Volmer for supply with the high temperature sample holder, Dr. A.K. Ariffin (Humboldt-Universität zu Berlin, and Universiti Pendidikan Sultan Idris, Malaysia) for providing the UHV measurement chamber, and T. Blume for technical support of the XAS experiment.